\documentclass{egpubl}
\usepackage{egsr15-e}

\WsPaper         
\electronicVersion 

\ifpdf \usepackage[pdftex]{graphicx} \pdfcompresslevel=9
\else \usepackage[dvips]{graphicx} \fi

\PrintedOrElectronic

\usepackage{t1enc,dfadobe}

\usepackage{egweblnk}
\usepackage{cite}

\usepackage{listings}

\lstdefinelanguage{GLSL}
{
sensitive=true,
morekeywords=[1]{
attribute, const, uniform, varying,
layout, centroid, flat, smooth,
noperspective, break, continue, do,
for, while, switch, case, default, if,
else, in, out, inout, float, int, void,
bool, true, false, invariant, discard,
return, mat2, mat3, mat4, mat2x2, mat2x3,
mat2x4, mat3x2, mat3x3, mat3x4, mat4x2,
mat4x3, mat4x4, vec2, vec3, vec4, ivec2,
ivec3, ivec4, bvec2, bvec3, bvec4, uint,
uvec2, uvec3, uvec4, lowp, mediump, highp,
precision, sampler1D, sampler2D, sampler3D,
samplerCube, sampler1DShadow,
sampler2DShadow, samplerCubeShadow,
sampler1DArray, sampler2DArray,
sampler1DArrayShadow, sampler2DArrayShadow,
isampler1D, isampler2D, isampler3D,
isamplerCube, isampler1DArray,
isampler2DArray, usampler1D, usampler2D,
usampler3D, usamplerCube, usampler1DArray,
usampler2DArray, sampler2DRect,
sampler2DRectShadow, isampler2DRect,
usampler2DRect, samplerBuffer,
isamplerBuffer, usamplerBuffer, sampler2DMS,
isampler2DMS, usampler2DMS,
sampler2DMSArray, isampler2DMSArray,
usampler2DMSArray, struct},
morekeywords=[2]{
radians,degrees,sin,cos,tan,asin,acos,atan,
atan,sinh,cosh,tanh,asinh,acosh,atanh,pow,
exp,log,exp2,log2,sqrt,inversesqrt,abs,sign,
floor,trunc,round,roundEven,ceil,fract,mod,modf,
min,max,clamp,mix,step,smoothstep,isnan,isinf,
floatBitsToInt,floatBitsToUint,intBitsToFloat,
uintBitsToFloat,length,distance,dot,cross,
normalize,faceforward,reflect,refract,
matrixCompMult,outerProduct,transpose,
determinant,inverse,lessThan,lessThanEqual,
greaterThan,greaterThanEqual,equal,notEqual,
any,all,not,textureSize,texture,textureProj,
textureLod,textureOffset,texelFetch,
texelFetchOffset,textureProjOffset,
textureLodOffset,textureProjLod,
textureProjLodOffset,textureGrad,
textureGradOffset,textureProjGrad,
textureProjGradOffset,texture1D,texture1DProj,
texture1DProjLod,texture2D,texture2DProj,
texture2DLod,texture2DProjLod,texture3D,
texture3DProj,texture3DLod,texture3DProjLod,
textureCube,textureCubeLod,shadow1D,shadow2D,
shadow1DProj,shadow2DProj,shadow1DLod,
shadow2DLod,shadow1DProjLod,shadow2DProjLod,
dFdx,dFdy,fwidth,noise1,noise2,noise3,noise4,
EmitVertex,EndPrimitive},
morekeywords=[3]{
gl_VertexID,gl_InstanceID,gl_Position,
gl_PointSize,gl_ClipDistance,gl_PerVertex,
gl_Layer,gl_ClipVertex,gl_FragCoord,
gl_FrontFacing,gl_ClipDistance,gl_FragColor,
gl_FragData,gl_MaxDrawBuffers,gl_FragDepth,
gl_PointCoord,gl_PrimitiveID,
gl_MaxVertexAttribs,gl_MaxVertexUniformComponents,
gl_MaxVaryingFloats,gl_MaxVaryingComponents,
gl_MaxVertexOutputComponents,
gl_MaxGeometryInputComponents,
gl_MaxGeometryOutputComponents,
gl_MaxFragmentInputComponents,
gl_MaxVertexTextureImageUnits,
gl_MaxCombinedTextureImageUnits,
gl_MaxTextureImageUnits,
gl_MaxFragmentUniformComponents,
gl_MaxDrawBuffers,gl_MaxClipDistances,
gl_MaxGeometryTextureImageUnits,
gl_MaxGeometryOutputVertices,
gl_MaxGeometryOutputVertices,
gl_MaxGeometryTotalOutputComponents,
gl_MaxGeometryUniformComponents,
gl_MaxGeometryVaryingComponents,gl_DepthRange},
morecomment=[l]{//},
morecomment=[s]{/*}{*/},
morecomment=[l][keywordstyle4]{\#},
}
\usepackage{algorithm}
\usepackage{algpseudocode}

\usepackage[rgb]{xcolor}
\usepackage{pdfcomment}

\title[Implementing a Photorealistic Rendering System using GLSL]%
      {\vspace{-2.5EM}Implementing a Photorealistic Rendering System using GLSL\vspace{-1EM}}

\author[T. Hachisuka]{Toshiya Hachisuka\vspace{-0.5EM}\\The Unviersity of Tokyo}

\begin{document}
\teaser{\vspace{-2.5EM}	
	\centering
	\begin{tabular}{@{}c@{}c@{}c@{}}
		\includegraphics[height=0.36\linewidth]{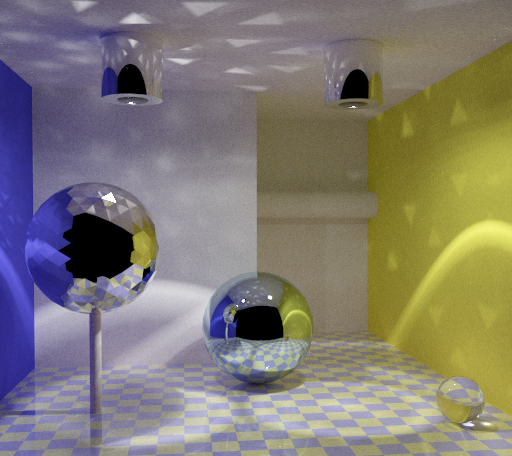} &
		\includegraphics[height=0.36\linewidth]{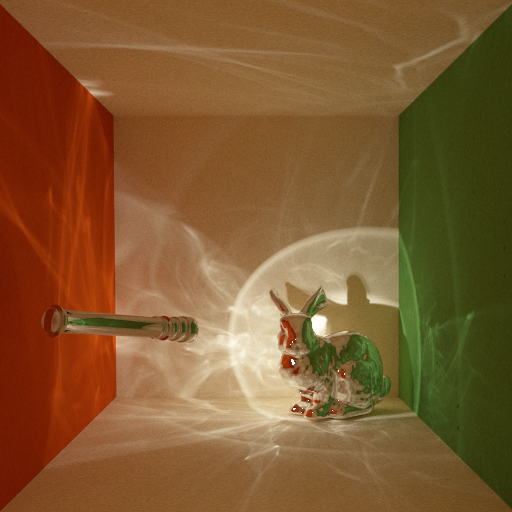} &
		\includegraphics[height=0.36\linewidth]{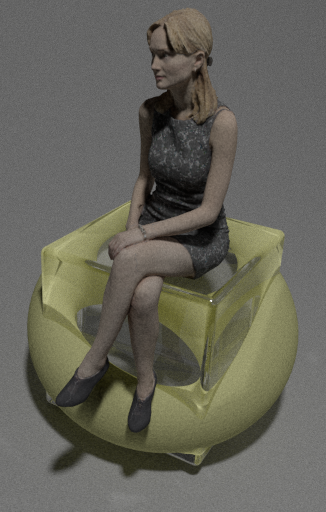} 
	\end{tabular}
	\caption{\label{fig:teaser}Example rendered images by the proposed system. All the images are rendered within one minute on GeForce~GTX~680. The rendering process runs entirely on a GPU using only GLSL shaders. The system runs equally fine on many different platforms while being capable of handling complex light transport paths.}
}

\maketitle

\begin{abstract}
Ray tracing on GPUs is becoming quite common these days. There are many publicly available documents on how to implement basic ray tracing on GPUs for spheres and implicit surfaces. We even have some general frameworks for ray tracing on GPUs. We however hardly find details on how to implement more complex ray tracing algorithms themselves that are commonly used for photorealistic rendering. This paper explains an implementation of a stand-alone rendering system on GPUs which supports the bounding volume hierarchy and stochastic progressive photon mapping. The key characteristic of the system is that it uses only GLSL shaders without relying on any platform dependent feature. The system can thus run on many platforms that support OpenGL, making photorealistic rendering on GPUs widely accessible. This paper also sketches practical ideas for stackless traversal and pseudorandom number generation which both fit well with the limited system configuration. 
\end{abstract}

\section{Introduction}
The use of GPUs for ray tracing is becoming increasingly common. In particular, details on how to implement a simple ray tracing system for a small number of objects or implicit surfaces are well documented in many publicly available tutorials. On the other hand, while there are several publicly available rendering systems on GPUs, implementation of such a more practical rendering system has been rarely documented. Nvidia's OptiX~\cite{parker2010optix} is one exception, yet OptiX itself is merely a general framework where we can implement various ray tracing algorithms on top of it. Even with the availability of such a general ray tracing framework, details on implementations of specific ray tracing algorithms have to be sorted out.

This paper explains an implementation of a rendering system which supports the bounding volume hierarchy~\cite{wald2007fast} and stochastic progressive photon mapping~\cite{hachisuka2009stochastic} using only OpenGL~3.0 and GLSL~1.20. This limited system configuration was chosen for multiple practical reasons. Firstly, a program can reliably run on various operating systems and GPUs. While this platform independence of OpenGL is not perfect, it is mostly true for battle-tested versions such as the version~3.0. This is in contrast to vendor-specific implementations~\cite{parker2010optix,davidovivc2014progressive} which are bound to be incompatible with GPUs of other vendors. Secondly, parallelization over multiple GPUs is automatically supported without additional code. Unlike OpenCL, a graphics driver manages multiple GPUs with a general parallelization strategy. While this strategy can be suboptimal for each application, this separation of management simplifies the implementation. Thirdly, a program can potentially run on web browsers via WebGL, since WebGL is essentially a limited version of OpenGL and becoming rapidly common. While WebCL~\cite{webcl} proposes support for more general GPU computation on any compatible web browser, no browsers currently support WebCL. There is no one to one correspondence between OpenGL and WebGL, but the proposed system uses mainly the features that are also available on WebGL. It is thus still making sense and practical to consider developing a rendering system using a rather old version of OpenGL for general computation. 

This paper specifically sketches two practical ideas which are suitable for this limited system configuration. The first idea is a modification of the threaded bounding volume hierarchy~\cite{simonsen2005comparison} which improves the traversal performance by two to three times. The modification is to pre-sort all the nodes in a given threaded BVH along principal directions of rays and to store multiple instances of threaded BVHs. The modified traversal algorithm remains simple and its performance is on a practical level. The second idea is a pseudorandom number generator which uses only floating-point numbers. This generator is computationally inexpensive while the quality of random numbers is sufficient. To summarize, the contributions are:

\begin{itemize}
	\item Open source GPU rendering system which uses limited features of OpenGL and thus is likely to run on WebGL.
	\item Modification of the threaded BVH which allows an efficient and simple stackless traversal of a given BVH.
	\item Introduction of a pseudorandom number generator which uses only floating-point number operations.
\end{itemize}

\begin{figure}[t]
	\includegraphics[width=\columnwidth]{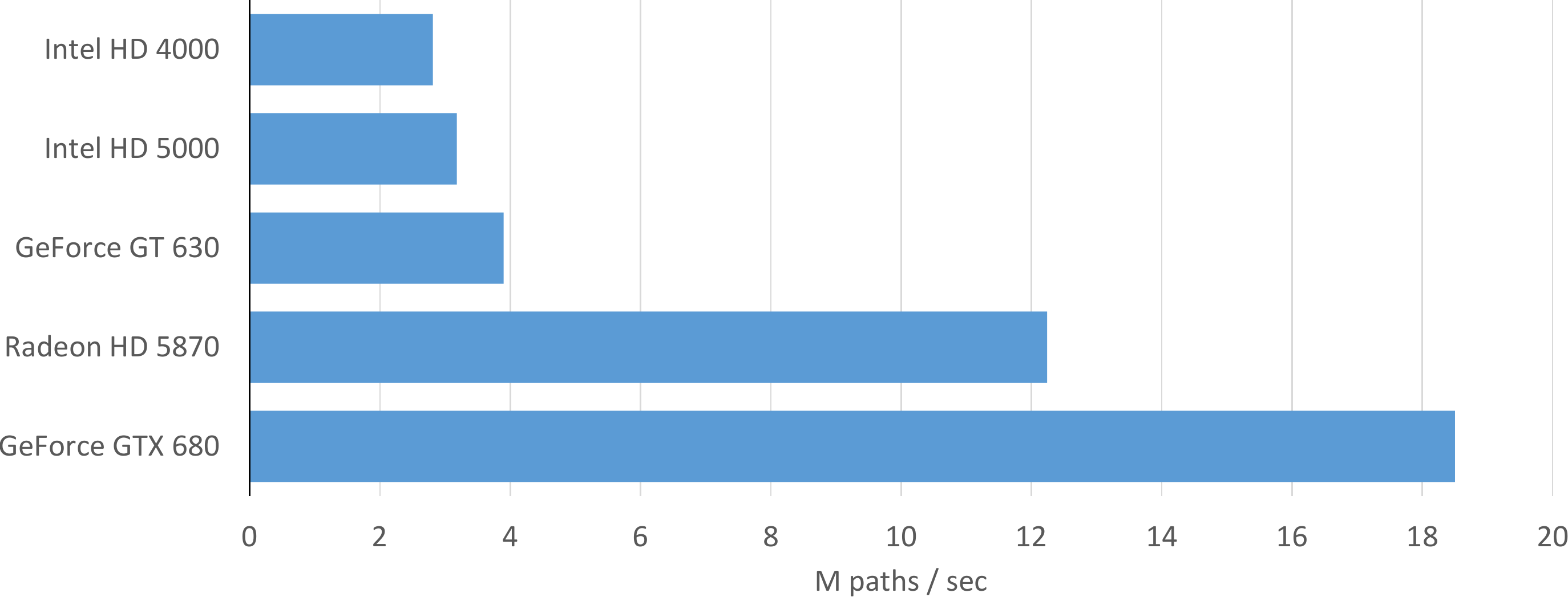}
\vspace{-0.6cm}
\caption{\label{fig:end}End-to-end performance on various GPUs on the metal Cornell box scene. The scene configuration is used by default in the released code.\vspace{-0.6cm}}
\end{figure}

While similar work has been recently published by Davidovi\v{c} et al.~\cite{davidovivc2014progressive}, their work focuses on a highly optimized implementation on Nvidia's GPUs using CUDA~\cite{CUDA}. The proposed system, on the other hand, is designed to be vendor independent. Since explaining all the details is not very informative, the following sections outline only some high-level ideas. For more details, please refer to the released code. Figure~\ref{fig:end} shows end-to-end performance on various GPUs.
The code is available at \url{http://www.ci.i.u-tokyo.ac.jp/~hachisuka/glslppm.zip} as of May 2015.
\vspace{-1EM}
\section{Overview}
The proposed system supports stochastic progressive photon mapping~\cite{hachisuka2009stochastic} as the main rendering engine. Each iteration of stochastic progressive photon mapping consists of three stages. The first stage is photon tracing which samples light paths starting from light sources. The second stage traces eye paths from the camera until they hit non-specular surfaces. The last stage performs range queries at the intersection points of eye paths and executes stochastic progressive density estimation. The overall implementation is not done by merely porting existing algorithms to GLSL, but comes with several algorithmic modifications as noted later. 

Both the photon tracing stage and eye ray tracing stage need an efficient ray casting algorithm. While there exist many efficient ray casting algorithms~\cite{Aila:Efficiency:NVIDIA:2012}, many of them are not compatible with our limited system configuration. The proposed system thus employs the threaded BVH~\cite{simonsen2005comparison} (TBVH) to achieve a simple stackless traversal algorithm. The original algorithm unfortunately has major performance degradation due to the fixed traversal order. The proposed modification, the multiple-threaded BVH (MTBVH), alleviates this issue and improves the traversal performance by two to three times. 

Monte Carlo sampling of light paths and eye paths needs an efficient and high quality random number generator. The main difficulty is that our limited system configuration does not allow any native bitwise operations. Existing random number generators on GPUs rely on the availability of bitwise operations~\cite{tzeng2008parallel} and thus are incompatible with our configuration. We introduce a random number generator that uses only floating-point number operations.

\vspace{-1EM}
\section{System components}
\subsection{Multiple-Threaded BVH}
Since GLSL~1.20 (and GLSL in WebGL) does not support stack, we need a stackless traversal algorithm that runs efficiently on GPUs. On top of this practical reason, even if we could use stack, stackless traversal is known to have multiple advantages for massively parallel platforms~\cite{afra2014stackless} such as GPUs. Threading~\cite{simonsen2005comparison} is one such approach to achieve stackless traversal of a given BVH. The threaded BVH stores hit/miss links instead of the tree structure to represent a given BVH. The traversal algorithm simply follows a hit or miss link depending on the result of the ray-AABB intersection test at each node.

\begin{figure}[t]
	\includegraphics[width=\columnwidth]{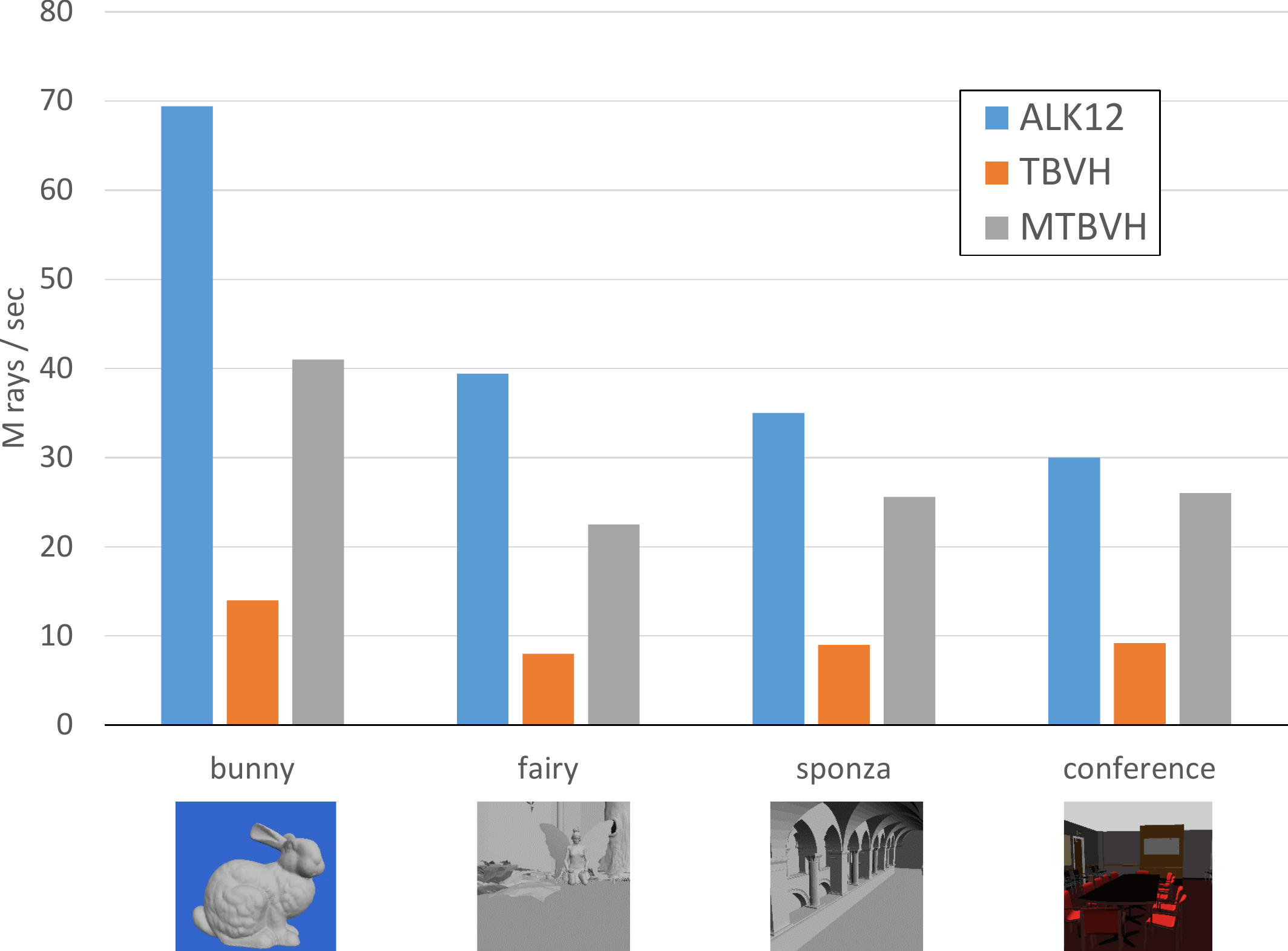}
\vspace{-0.2cm}
\caption{\label{fig:perf}Performance comparisons of ray casting using the vendor-specific optimized traversal~\protect\cite{Aila:Efficiency:NVIDIA:2012}, the original threaded BVH~\protect\cite{simonsen2005comparison} (TBVH), and the multiple-threaded BVH (MTBVH). The experiments used the SAH-BVH~\protect\cite{wald2007fast} for all the algorithms and the computation times include diffuse shading. The testing environment is GeForce GT 630. \vspace{-20pt}}
\end{figure}

The proposed modification is to store six instances of hit/miss links by pre-sorting nodes along positive x axis, negative x axis, and so on for y and z as well. The traversal algorithm remains almost the same, but it now selects one out of the six sets of links at the beginning, depending on the direction of a given ray. Figure~\ref{fig:mtbvh} is the pseudocode of the traversal algorithm with the highlighted modification. This simple modification enables approximate optimization of the traversal order according to a given ray direction. Figure~\ref{fig:perf} shows comparisons of ray traversal performance. While the multiple-threaded BVH is still not as fast as the vendor-specific optimization~\cite{Aila:Efficiency:NVIDIA:2012}, it is two to three times faster than original threaded BVH and remains vendor-independent. It should be emphasized that the proposed algorithm is not fundamentally designed to achieve the best performance, but to achieve reasonable performance with only vendor-independent features. Please be aware that the provided code shows the number of complete paths per second including end-to-end rendering computation, not raw rays per second, thus the numbers will be different from those in Figure~\ref{fig:perf}.

\begin{figure}[t]
\tiny
\begin{lstlisting}[language=GLSL]
<@\textcolor{red}{node = cubemap(root\_tex, ray.direction)}@>;
while (node != null) {
	if (intersect(node.aabb, ray)) {
		if (node.leaf) result = intersect(node.triangles, ray);
		node = node.hit;
	} else {
		node = node.miss;
	}
}
\end{lstlisting}
\vspace{-0.2cm}
\caption{\label{fig:mtbvh}Traversal algorithm of MTBVH. The only difference from the traversal algorithm of the original threaded BVH~\protect\cite{simonsen2005comparison} is that it chooses an pre-ordered set of hit/miss links according to the ray direction (first line).\vspace{-0.4cm}}
\end{figure}


The implementation uses a standard top-down sweeping algorithm for constructing a BVH based on SAH~\cite{wald2007fast}. The construction and threading are both currently done on CPUs and the resulting data is transferred to GPUs afterward. Threading is usually done less than 100~ms even for a typical scene and hardly becomes the bottleneck. 

The storage overhead of MTBVH is not as significant as it appears to be. For instance, if we count the number of \textsc{vec4}s used in original threaded BVH, a triangle is stored as six \textsc{vec4}s (two \textsc{vec4}s for the packed position, normal, and texture coordinates for each vertex), an AABB is stored as two \textsc{vec4}s (min and max), and hit/miss links can be packed into one \textsc{vec4}. MTBVH adds only five more hit/miss links. Since the number of triangles and the number of nodes are typically not very different, MTBVH does not increase the total storage cost by six times, but by approximately 1.56 times (9 \textsc{vec4}s of the original vs 14 \textsc{vec4}s of ours). Since image textures usually add to the storage cost significantly more, the overhead of MTBVH is not significant.

%
%
%


\vspace{-1EM}
\subsection{PRNG using only floating point numbers}
Since GLSL~1.20 does not support bitwise arithmetic operations, we cannot port existing pseudorandom number generators (PRNGs) such as the one based on cryptographic hashing~\cite{tzeng2008parallel}. The system thus uses a weighted sum of multiplicative linear congruential generators~\cite{l1988efficient} only with floating-point number operations. 
Figure~\ref{fig:rand} shows the GLSL code of this PRNG. The original version of the algorithm was introduced as an anonymous post at the GPGPU web forum (\url{http://web.archive.org/web/20101217080108/http://gpgpu.org/forums/viewtopic.php?t=2591}). The algorithm in Figure~\ref{fig:rand} has been modified to run well on GLSL. To generate many random numbers in parallel, one can initialize each PRNG state on a CPU via xorshift~\cite{marsaglia2003xorshift}. 
\begin{figure}[t]
\tiny
\begin{lstlisting}[language=GLSL]
float GPURnd(inout vec4 state) {
  const vec4 q = vec4(1225, 1585, 2457, 2098);
  const vec4 r = vec4(1112, 367, 92, 265);
  const vec4 a = vec4(3423, 2646, 1707, 1999);
  const vec4 m = vec4(4194287, 4194277, 4194191, 4194167);
	
  vec4 beta = floor(state / q);
  vec4 p = a * mod(state, q) - beta * r;
  beta = (sign(-p) + vec4(1)) * vec4(0.5) * m;
  state = (p + beta);
	
  return fract(dot(state / m, vec4(1, -1, 1, -1)));
}
\end{lstlisting}
\vspace{-0.2cm}
\caption{\label{fig:rand}Pseudorandom number generator using only floating point number operations. The algorithm is based on a weighted sum of four instances of the multiplicative linear congruential generator~\protect\cite{l1988efficient}.\vspace{-0.8cm}}
\end{figure}

Note that the original algorithm~\cite{l1988efficient} uses integer arithmetic. While the authors have not observed any major artifacts due this change, changing the algorithm to use only floating point numbers might have introduced some statistical deficiency. The author of the above anonymous posting seems to claim certain quality of randomness within the 23 bit mantissa. 

\vspace{-1EM}
\subsection{Photon path regeneration}
A standard implementation of photon tracing keeps a global list of photons and sequentially adds a photon to the list. This approach however needs inter-threads commutation on GPUs to compact lists in order to obtain a complete global list of photons. The alternative approach used in the implementation is to trace a single bounce per pass and stores only one photon at most. For example, the very first pass traces photons from light sources toward the first intersections and stores the photons. Further bounces are traced only in succeeding passes. Each pass thus outputs at most one photon per thread, not a list of photons.

A new photon path is generated at next pass if the current photon path is killed by Russian Roulette or misses a scene. This process is done independently between threads (in our case, threads are equal to pixels), thus each thread potentially traces a photon ray at a different number of bounces. This approach keeps all threads busy all the time regardless of path length. A similar method is used for path tracing~\cite{Novak:2010:PathReg} and they reported performance improvement. We can observe similar improvement in the proposed system.


\vspace{-1EM}
\subsection{Stochastic spatial hashing}
Hachisuka and Jensen~\cite{hachisuka2010parallel} proposed a stochastic hashing algorithm that utilizes the statistical nature of density estimation. They pointed two fundamental challenges in the use of regular spatial hashing on GPUs: the sequential nature of list constructions for hash table entries and uneven workload distribution at the data retrieval phase. Their key idea is to let only one data survive for each hash table entry with concurrent writes, and to scale the contribution of each photon by the number of hash collisions. 

The proposed modification is to assign a statistically independent random depth value (using the PRNG in Figure~\ref{fig:rand}) for each photon path and to enable z-buffering for hashing photon data simultaneously. Davidovi\v{c} et al.~\cite{davidovivc2014progressive} pointed that stochastic hashing can potentially produce wrong results if the timing that each photon data is hashed depends on some properties of the given photon path such as path length. This modification ensures that the probability that one photon survives over other photons is independent of any properties of the corresponding photon path. 

\vspace{-1EM}
\section{Discussion}
All the shaders in the proposed system use features only of GLSL~1.20. They can thus run on WebGL which is based on OpenGL~ES~2.0 and supports most features of GLSL~1.20. In practice, however, they do not run on WebGL alone, since the official specification of WebGL does not guarantee a native support of some important features such as rendering to floating point number textures. It should also be repeated that there is no one to one correspondence between a certain version of OpenGL and WebGL. Having said that, since WebGL can support floating point textures via the extension, it is most likely possible to port the code for WebGL. WebGL 2~\cite{webgl2} will natively support this extension.

\subsection{Limitations and future work}
While the code is intended to be platform-independent, it is possible to fail due to the vendor-dependent JIT compilation model of GLSL. No fundamental modification, however, will be necessary to make the code successfully executable. The author would appreciate a bug report in case you found any. 
The current implementation does not build an acceleration data structure using GPUs. It is however trivial to implement linear BVH~\cite{lauterbach2009fast} using bitonic sorting~\cite{purcell2003photon} within our limited configuration, if one wants to build an acceleration data structure on GPUs. A more efficient construction for a high quality tree~\cite{garanzha2011simpler} using only GLSL could however be challenging due the requirement of a more advanced thread management such as work queue.
While stochastic progressive photon mapping covers many scene configurations, one might want to implement more advanced rendering algorithms such as unified path sampling~\cite{hachisuka2012path}~/~vertex merging and connection~\cite{georgiev2012light}. Davidovi\v{c} et al.~\cite{davidovivc2014progressive} show how to implement some of such algorithms using CUDA, but porting their algorithms on GLSL may pose some challenges.
The proposed system does not support out-of-core rendering. An efficient out-of-core rendering on GPUs is still an open problem even with more general GPU computation platforms. This feature, however, might not be necessary for some applications such as rendering for online shopping.
\section{Conclusion}
The paper outlines an implementation of a rendering system via OpenGL~3.0 and GLSL~1.20 without relying on any platform dependent feature. One of the proposed modifications, the {multiple-threaded BVH}, uses the principal direction of a given ray to select one of the threaded BVHs. This simple modification results in two to three times performance improvement compared to the original threaded BVH. Unlike common pseudorandom number generators, the proposed generator uses only floating point number operations based on a combination of multiplicative linear congruential generators. The resulting generator is computationally inexpensive and the quality of generated random numbers is enough for the proposed rendering system. 
%
The author believes that the code can serve as an example implementation of a rendering system using only platform independent features. 

\paragraph*{Acknowledgements}
Thank Kentaro Oku (kioku@sys-k.net) for reporting bugs of the first release of the code. The scanned woman model is by courtesy of FUTURESCAN.

\bibliographystyle{eg-alpha-doi}

\bibliography{egbibsample}

%
%

\end{document}